\documentclass[11pt]{article}

\pdfoutput=1
\usepackage{arxiv}
\usepackage{hyperref}
\usepackage[normalem]{ulem}
\usepackage{amsthm}
\usepackage{amsmath}
\usepackage{amssymb}
\usepackage{mathrsfs}
\usepackage{mathtools}
\usepackage{graphicx}
\usepackage{enumerate}
\usepackage{comment}
\usepackage{algorithm}
\usepackage{algpseudocode}
\usepackage{color}
\usepackage{stmaryrd}
\usepackage{booktabs} 
\usepackage{fullpage}
\usepackage{appendix}
\usepackage[table]{xcolor}

\newtheorem{theorem}{Theorem}

\theoremstyle{definition}
 
\theoremstyle{remark}

\theoremstyle{remark}
\newtheorem{example}[theorem]{Example}
\numberwithin{theorem}{section}

\newcommand{\E}[1]{{\mathbb E}\left[#1\right]}

\newcommand\cC{\mathcal C}

\newcommand\cX{{\mathcal X}}
\newcommand\cY{{\mathcal Y}}

\newcommand{\vol}[1]{\ensuremath{\mathrm{vol}\left(#1\right)}}

\newcommand{\scorelabel}{\ensuremath{\mathrm{CAS}}}

\title{
Improving community detection via community association strength scores
}

\author{
Jordan Barrett\thanks{Department of Mathematics, Toronto Metropolitan University, Toronto, Canada; e-mail: \texttt{jordan.barrett@torontomu.ca}}
\And
Ryan DeWolfe\thanks{Department of Mathematics, Toronto Metropolitan University, Toronto, Canada; e-mail: \texttt{ryan.dewolfe@torontomu.ca}}
\And
Bogumi\l{} Kami\'{n}ski\thanks{Decision Analysis and Support Unit, SGH Warsaw School of Economics, Warsaw, Poland; email: \texttt{bkamins@sgh.waw.pl}}
\And
Pawe\l{} Pra\l{}at\thanks{Department of Mathematics, Toronto Metropolitan University, Toronto, Canada; e-mail: \texttt{pralat@torontomu.ca}}
\And
Aaron Smith\thanks{Department of Mathematics and Statistics, University of Ottawa, Ottawa, Canada; e-mail: \texttt{asmi28@uOttawa.ca}}
\And
Fran\c{c}ois Th\'{e}berge\thanks{Tutte Institute for Mathematics and Computing, Ottawa, Canada; email: \texttt{theberge@ieee.org}}
}

\begin{document}

\maketitle

\begin{abstract}
Community detection methods play a central role in understanding complex networks by revealing highly connected subsets of entities. However, most community detection algorithms generate partitions of the nodes, thus (i) forcing every node to be part of a community and (ii) ignoring the possibility that some nodes may be part of multiple communities. In our work, we investigate three simple community association strength (CAS) scores and their usefulness as post-processing tools given some partition of the nodes. We show that these measures can be used to improve node partitions, detect outlier nodes (not part of any community), and help find nodes with multiple community memberships. 
\end{abstract}

\section{Introduction}\label{sec:intro}

Detecting and analyzing the community structure in graphs is a fundamental problem in applied graph theory. Algorithms such as Girvan-Newman~\cite{newman2004finding}, Louvain~\cite{blondel2008fast}, and Leiden~\cite{traag2019louvain} attempt to partition the nodes of a graph into clusters with the general goal of maximizing edge density within clusters and minimizing edge density between clusters. These global partitioning algorithms have been used to detect community structure in various real-world networks such as collaboration networks~\cite{newman2004finding,luzar2014community}, biological networks~\cite{rahiminejad2019topological,calderer2021biological}, and social media networks~\cite{han2024social,Papadopoulos2012CommunityDI}, and have been tested and studied on synthetic models with ground-truth communities such as ABCD~\cite{kaminski2022modularity} and LFR~\cite{lancichinetti2008benchmark}. Typically, these community detection algorithms return a partition, meaning each node must be part of exactly one community. In many real-world networks, though, members can belong to 0, 1, or multiple communities. Some algorithms to detect overlapping communities include clique-percolation~\cite{derenyi2005clique}, edge clustering~\cite{ahn2010link} and ego-split~\cite{epasto2017ego}. A post-processing algorithm was proposed~\cite{jakatdar2022aoc} to detect overlapping communities, but this approach is based on $k$-core clustering and often only covers a small part of the nodes. This paper proposes post-processing a given node partition using local community aware scores called \textit{community association strength} (CAS) that are inspired by the measures proposed in~\cite{kaminski2024predicting}.

The rest of the paper is organized as follows.
In Section~\ref{sec:CAS}, we outline three CAS scores and show their similarities, differences, and relative abilities to predict community involvement. 
In Section~\ref{sec:post-processing}, we show how such CAS scores can be used for three different but complementary tasks: (i) improving an existing graph partitioning algorithm, (ii) detecting nodes not part of any community (outlier nodes), and (iii) detect multi-community node memberships. Finally, some concluding remarks are given in Section~\ref{sec:conclusion}

\section{Community association strength}\label{sec:CAS}

\subsection{CAS scores}\label{subsec:scores}

Given a graph $G = (V, E)$, a \scorelabel{} score is any function $f: V \times 2^V \to [0, 1]$. In practice, we aim to find a \scorelabel{} score $f$ such that $f(v, C)$ indicates how well $v$ is associated to community $C$, i.e., $f(v, C) \approx 1$ should imply that $v$ is strongly associated to community $C$ whereas $f(v, C) \approx 0$ should imply that $v$ has little to no association to community $C$. We will motivate and outline three such functions. 

\subsubsection*{Internal edge fraction $(\mathrm{IEF})$}

Arguably, the simplest measure for how strongly $v$ is associated with community $C$ is the \textit{internal edge fraction}. For a community $C \subseteq V$, write $\deg_C(v) := \left| \{(u, v) \in E : u \in C\} \right|$ and $\deg(v) := \deg_V(v)$. Then, the internal edge fraction of $v$ in $C$ is defined as
\[
\mathrm{IEF}(v, C) := \frac{\deg_C(v)}{\deg(v)} \,.
\]
Although the internal edge fraction is both easy to interpret and a strong indicator of community association, it fails to account for the size of the communities. For example, if $G$ contains a large community $C_1$ with, say, $|C_1| \approx |V|/2$ and a small community $C_2$ with $|C_2| \ll |V|$, then we may want to distinguish the outcomes $\mathrm{IEF}(v, C_1) = 1$ and $\mathrm{IEF}(v, C_2) = 1$, especially if $\deg(v)$ is large.

\subsubsection*{Normalized internal edge fraction $(\mathrm{NIEF})$}

For a graph $G = (V, E)$ and a community $C$, write $w(C) := {\vol{C}}/{\vol{V}}$ where $\vol{C} := \sum_{v \in C} \deg(v)$. The second \scorelabel{} score we consider is the \textit{normalized internal edge fraction}, defined as
\[
\mathrm{NIEF}(v, C) := \max\left\{ \mathrm{IEF}(v, C) - w(C) , 0 \right\} \,.
\]
This \scorelabel{} score is derived as follows. Let $G = (V, E)$, let $\mathbf{d} = \{\deg(v), v \in V\}$ and let $\widehat{G} \sim \mathrm{ChungLu}(\mathbf{d})$~\cite{chung2006complex}. Then, for any $v \in V$,
\[
\mathrm{NIEF}(v, C) := \max\left\{ \mathrm{IEF}_G(v, C) - \E{\mathrm{IEF}_{\widehat{G}}(v, C)}, 0 \right\} \,,
\]
where we assume that $C$ is fixed before sampling $\widehat{G}$.
In other words, $\mathrm{NIEF}(v, C)$ compares the actual internal edge fraction of $v$ to the ``expected'' such fraction under the associated null-model. Note that this score is inspired by the \textit{community association strength} feature presented in~\cite{kaminski2024predicting}. 

Consider again the previous example with $G$ containing communities $C_1$ and $C_2$ with $|C_1| \approx |V|/2$ and $|C_2| \ll |V|$. If $\mathrm{IEF}(v, C_1) = 1$ and $\mathrm{IEF}(v, C_2) = 1$ then we get that $\mathrm{NIEF}(v, C_1) \approx 0.5$ and $\mathrm{NIEF}(v, C_2) \approx 1$. These scores agree with the intuition that $v$ having all its connections into a small community is more surprising than into a large community. 

\subsubsection*{$\mathrm{P}$ score}
Finally, we consider a score based on the classic $p$-value significance test. Write $F(\cdot; n, p)$ for the CDF of the distribution $\mathrm{Binomial}(n, p)$. The $\mathrm{P}$ score is defined as
\[
\mathrm{P}(v, C) := F\left(\deg_C(v) - 1; \deg(v), w(C) \right) \,.
\]
Here, $1 - \mathrm{P}(v, C)$ is the probability that at least $\deg_C(v)$ edges join $v$ and $C$ in a Chung-Lu resampling of $G$ (keeping $C$ fixed), i.e., $1 - \mathrm{P}(v, C)$ is the classic $p$-value of $\deg_C(v)$ when considering the Chung-Lu model as the null model. Similar to the $\mathrm{NIEF}$ score, the $\mathrm{P}$ score is sensitive to the size of the community. Unlike $\mathrm{NIEF}$, however, the $\mathrm{P}$ score is also sensitive to the degree of the node. For example, in general we have that
\[
F(2\deg_C(v)-1; 2\deg(v), p) \neq F(\deg_C(v)-1; \deg(v), p) \,,
\] 
whereas it is always true that
\[
\frac{2\deg_C(v)}{2\deg(v)} - w(C) = \frac{\deg_C(v)}{\deg(v)} - w(C) \,.
\]
The idea here is that, keeping everything else equal, it is more ``surprising'' to see a high-degree node having a lot of edges in a community than for a low-degree node (which could happen by pure chance). Note that the P score is similar to, albeit distinct from, the community fitness measure used in~\cite{lancichinetti2011statisticallysignificantcommunities}.

\subsection{Comparing CAS scores}

All three \scorelabel{} scores, $\mathrm{IEF}$, $\mathrm{NIEF}$, and $\mathrm{P}$, share two qualitative properties that we believe are essential for indicating the association strength of a node into a community. Firstly, all three scores evaluate $0$ if no edges join $v$ and community $C$. Secondly, conditioned on $\vol{C}$, all three scores are monotone with respect to $\deg_C(v)$; if $\vol{C_1} = \vol{C_2}$ and $\deg_{C_1}(v) > \deg_{C_2}(v)$ then $(v, C_1)$ yields a higher \scorelabel{} score than $(v, C_2)$ under all three measures. 

Perhaps more interesting than the similarities between these three \scorelabel{} scores are their differences. As mentioned previously, a key difference between these scores is their sensitivity to the community size and the degree of the node: $\mathrm{P}(v, C)$ is sensitive to both $\deg(v)$ and $|C|$, $\mathrm{NIEF}(v, C)$ is sensitive only to $|C|$, and $\mathrm{IEF}$ is sensitive to neither. We will illustrate this point in the following two examples. 

\begin{example}\label{ex:sensitivity to |C|}
    Let $G = (V, E)$ be a graph with $\vol{V} = 10,000$, let $C_1, C_2 \subset V$ be communities in $G$ with $\vol{C_1} = \vol{V}/2 = 5,000$ and $\vol{C_2} = 100$, and let $v \in V$ be a node with $\deg(v) = 5$ such that all edges from $v$ connect to either $C_1$ or $C_2$. Table~\ref{tab:example1} shows the three \scorelabel{} scores as $\deg_{C_1}(v)$ and $\deg_{C_2}(v)$ vary, highlighting the point in each score where the association strength of $v$ into $C_2$ becomes larger than that of $C_1$. 
    
    We find that $\mathrm{IEF}(v, C_1) < \mathrm{IEF}(v, C_2)$ precisely when $\deg_{C_1}(v) < \deg_{C_2}(v)$. However, for both $\mathrm{NIEF}$ and $\mathrm{P}$, there is a large penalty associated with $C_1$ compared to a negligible penalty associated with $C_2$. Note that in this case, the bias towards the smaller community is stronger for $\mathrm{P}$ than for $\mathrm{NIEF}$. In our testing, we found this trend to be true in general.
\end{example} 

\begin{table}
    \[
    \begin{array}{|cc|cc|cc|cc|}
        \hline
        \deg_{C_1} & \deg_{C_2} & \mathrm{IEF}(C_1) & \mathrm{IEF}(C_2) & \mathrm{NIEF}(C_1) & \mathrm{NIEF}(C_2) & \mathrm{P}(C_1) & \mathrm{P}(C_2) \\
        \hline \hline
        5 & 0 & 1 & 0 & 0.5 & 0 & 0.97 & 0 \\
        \hline
        4 & 1 & 0.8 & 0.2 & 0.3 & 0.19 & \cellcolor{gray!25}0.81 & \cellcolor{gray!25}0.95 \\
        \hline
        3 & 2 & 0.6 & 0.4 & \cellcolor{gray!25}0.1 & \cellcolor{gray!25}0.39 & \cellcolor{gray!25}0.5 & \cellcolor{gray!25}1 \\
        \hline
        2 & 3 & \cellcolor{gray!25}0.4 & \cellcolor{gray!25}0.6 & \cellcolor{gray!25}0 & \cellcolor{gray!25}0.59 & \cellcolor{gray!25}0.19 & \cellcolor{gray!25}1 \\
        \hline
        1 & 4 & \cellcolor{gray!25}0.2 & \cellcolor{gray!25}0.8 & \cellcolor{gray!25}0 & \cellcolor{gray!25}0.79 & \cellcolor{gray!25}0.03 & \cellcolor{gray!25}1 \\
        \hline
        0 & 5 & \cellcolor{gray!25}0.0 & \cellcolor{gray!25}1.0 & \cellcolor{gray!25}0 & \cellcolor{gray!25}0.99 & \cellcolor{gray!25}0 & \cellcolor{gray!25}1 \\
        \hline
    \end{array}
    \]
    \caption{A comparison of CAS scores for 2 communities $C_1, C_2 \subset V$ with $\vol{V} = 10,000$, $\vol{C_1} = 5,000$ and $\vol{C_2} = 100$. The scores are rounded to 2 decimal places, and node $v$ is omitted from the notation. Here, $\deg(v) = 5$ and each row represents a different split of $\deg(v)$ into $C_1$ and $C_2$. The grey cells highlight when a CAS score favours $C_2$ over $C_1$.}
    \label{tab:example1}
\end{table}

\begin{example}
    Consider a graph $G = (V, E)$, a community $C$ with $\vol{C} = \vol{V}/2$, and a node $v \in V$. Table~\ref{tab:example2} shows $\mathrm{NIEF}$ and $\mathrm{P}$ scores for $(v, C)$ as $\deg(v)$ and $\deg_C(v)$ vary whilst $\deg_C(v)/\deg(v)$ remains constant. 
    
    The increasing values of $\mathrm{P}(v, C)$ as $\deg(v)$ and $\deg_C(v)$ increase is a consequence of the law of large numbers for the binomial distribution. Ultimately, what this example shows is that the $\mathrm{P}$ score, conditioned on $\deg_C(v) / \deg(v) > w(C)$, is biased towards nodes with a larger degree, whereas the $\mathrm{NIEF}$ score has no such bias. 
\end{example}

\begin{table}
    \[
    \begin{array}{|cc|c|c|}
        \hline
        ~\deg~ & ~\deg_C~ & ~\mathrm{NIEF}(C)~ & ~\mathrm{P}(C)~ \\
        \hline \hline
        3 & 2 & 0.17 & 0.5 \\
        \hline
        6 & 4 & 0.17 & 0.66 \\
        \hline
        9 & 6 & 0.17 & 0.75 \\
        \hline
        12 & 8 & 0.17 & 0.81 \\
        \hline
        15 & 10 & 0.17 & 0.85 \\
        \hline
    \end{array}
    \]
    \caption{The comparison of 2 CAS scores on community $C$ with $\vol{C} = \vol{V}/2$. The scores are rounded to 2 decimal places, and the node $v$ is omitted from the notation.}
    \label{tab:example2}
\end{table}

\subsection{The quality of CAS scores}
\label{subsec:CAS quality}

To test the quality of the various community association strength scores, we need graphs with ground-truth communities, where the communities do not necessarily form a partition of the nodes. To this end, we will test a family of synthetic models stemming from the Artificial Benchmark for Community Detection (ABCD) model. The ABCD model ~\cite{kaminski2021artificial} is a synthetic model with ground-truth communities and has 8 parameters to control the number of nodes, the degree and community size distributions, and the fraction $\xi$ of noise. A fast, multi-threaded implementation of ABCD (ABCDe) was introduced in~\cite{kaminski2022abcde}, and a hypergraph generalization (h-ABCD) was introduced in~\cite{kaminski2022hypergraph}. Importantly for our research, the ABCD model was generalized (ABCD+o) to include outliers~\cite{kaminski2023artificial}, and further generalized (ABCD+o$^2$) to allow for overlapping communities. The latter model includes a parameter $\eta \geq 1$ which governs the expected number of community memberships for non-outlier nodes. 

\begin{figure}[ht]
    \centering
    \includegraphics[width=8cm]{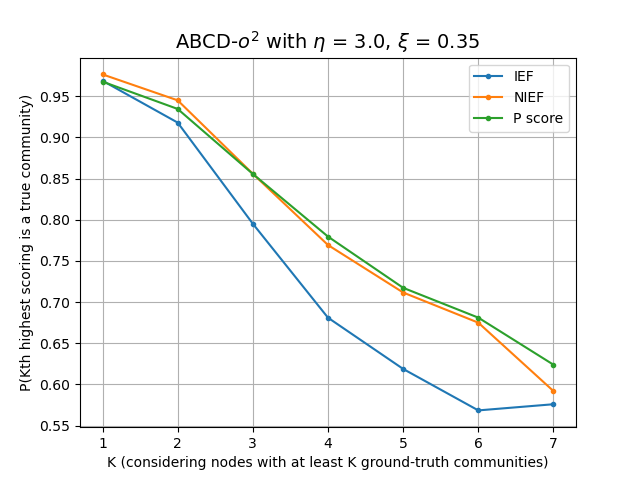}
    \includegraphics[width=8cm]{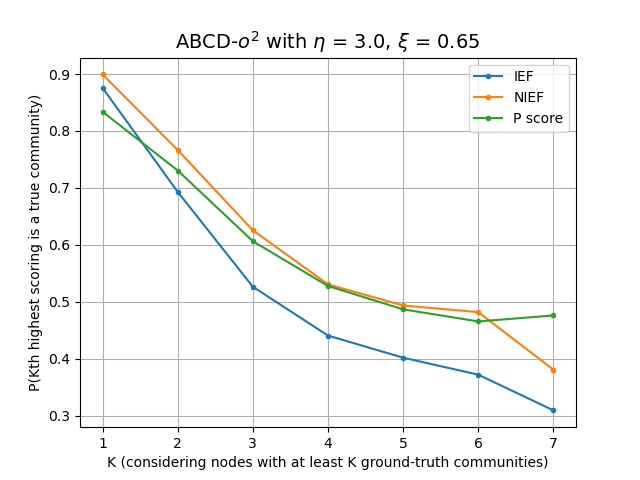}
    \caption{Proportion of $K^{th}$ highest scoring community that are actually ground-truth communities for the three CAS scores. Results are averaged over 10 ABCD+o$^2$ graphs with 10,000 nodes including 250 outliers, overlap parameter $\eta=3$ and noise parameters $\xi=0.35$ (left) and $\xi=0.65$ (right). We can see that with all scores, the first few highest-scoring communities are almost always ground-truth, and this slowly degrades as $K$ increases, with the NIEF and P scores decreasing more slowly. }
    \label{fig:k-rank-abcdoo}
\end{figure}

We first test the \scorelabel{} scores' ability to rank communities based on the likelihood of a node being in those communities. For a graph $G$, a collection of ground-truth communities $\cC$, and a \scorelabel{} score $f: V \times 2^V \to [0, 1]$, let $V_k \subseteq V$ be the set of nodes such that $v \in V_k$ if and only if $v$ is contained in at least $k$ communities in $\cC$. For each score $f$, each relevant $k$, and each $v \in V_k$, we consider the $k^{th}$ highest ranking community in $\cC$ according to $f$ and check if this community indeed contains $v$. Figure~\ref{fig:k-rank-abcdoo} presents the experiment results using ABCD+o$^2$ graphs with two different noise parameters: $\xi = 0.35$ and $\xi = 0.65$. All three measures perform similarly, with $\mathrm{NIEF}$ and $\mathrm{P}$ performing slightly better. The results suggest that each of the measures can accurately predict 1 or 2 communities a node is a member of, and with a low noise parameter, the prediction accuracy remains high as the number of communities increases. 

\begin{figure}[ht]
    \centering
    \includegraphics[width=8cm]{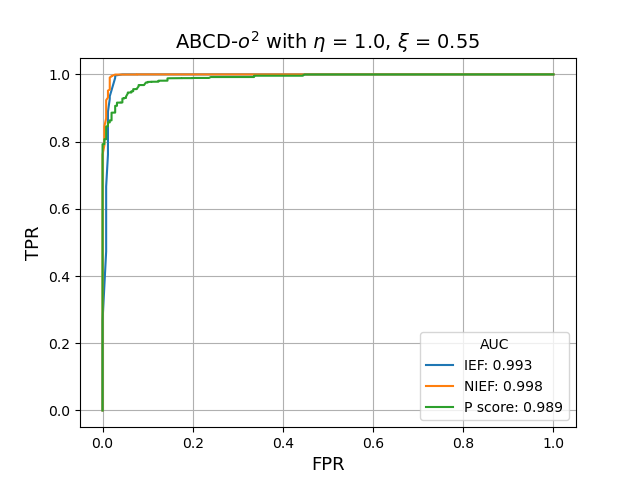}
    \includegraphics[width=8cm]{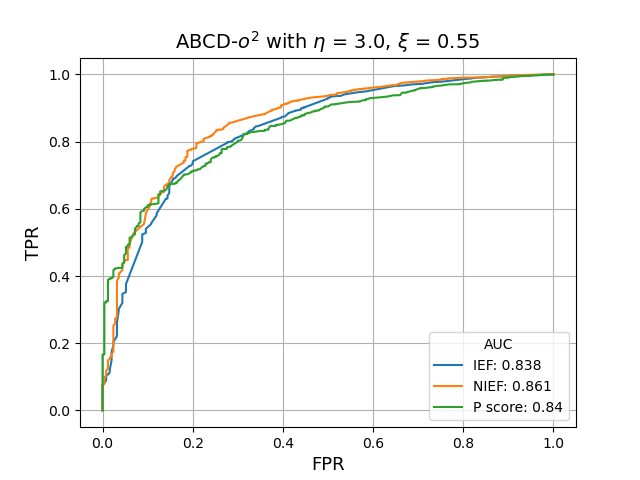}
    \caption{ROC curves and corresponding AUC measures for three CAS scores for ABCD+o$^2$ graphs.}
    \label{fig:abcdoo_roc}
\end{figure}

Next, we test the scores' ability to distinguish outliers from non-outliers. Let $V_o \subseteq V$ be the outlier nodes concerning ground-truth communities $\cC$. For each \scorelabel{} score $f$, we order the nodes in $V$ from smallest to largest based on $\max\{f(v, \cdot )\}$. We predict that, for $u, v \in V$ with $\max\{f(u, \cdot)\} < \max\{f(v, \cdot)\}$, $u$ is more likely to be an outlier than $v$. We then compare our prediction to the ground truth with a receiver operating characteristic (ROC) curve. Figure~\ref{fig:abcdoo_roc} summarizes the results of this experiment by showing ROC curves for ABCD+o$^2$ graphs with 10,000 nodes (including 250 outliers), moderate noise ($\xi=0.55$) and two different values for $\eta$. We also show each score's area under the ROC curves (AUC). While all measures get almost perfect results when $\eta=1$ (no community overlap), we see degradation in the presence of community overlap ($\eta=3$). We again see that all three measures perform similarly, with NIEF performing slightly better.

These two experiments indicate that CAS scores can capture local properties useful to refine node partitions. In the next section, we consider more realistic experiments where we do not explicitly use the ground-truth communities (but we use them for evaluation).

\section{Using CAS scores in practice}
\label{sec:post-processing}

In the previous section, we tested the quality of CAS scores with respect to the ground-truth communities of a graph. In practice, however, we may not have access to the ground-truth communities and would thus like to leverage CAS scores to help recover said communities. To this end, we present three scenarios where CAS scores can be useful. Section \ref{subsec:wECG} describes an improvement to an existing consensus clustering algorithm. We then show applications of CAS scores to post-process any partitioning algorithm: identifying outlier nodes in Section~\ref{subsec:outliers} and identifying multi-community memberships in Section~\ref{subsec:multi-community vetices}. Finally, we apply all three ideas to the well known college football graph in Section~\ref{subsec:football}.

\subsection{Improving graph partitions}
\label{subsec:wECG}

In this section, we use CAS scores to modify an existing clustering algorithm called Ensemble Clustering for Graphs (ECG)~\cite{poulin2018ensemble,poulin2019ecgcomparisons} which is itself a modification of the Louvain algorithm. Let us briefly describe the ECG, and then our modification to ECG (which we will refer to as CAS-ECG).

The ECG algorithm takes a positive integer $k$ as input and executes as follows. 
\begin{enumerate}
\item Perform the first iteration of the standard Louvain algorithm on $(V, E)$, independently $k$ times, to yield $k$ ``level 1'' partitions of $V$. 
\item Weight $E$ such that the weight of edge $(u, v)$ is proportional to the number partitions from step~1 with $u$ and $v$ in the same part.  
\item Run Louvain, Leiden or some other partitioning algorithm on this re-weighted graph. 
\end{enumerate}
The full description of the algorithm can be found in Section~3 of~\cite{poulin2018ensemble}.

We now detail CAS-ECG. During step~2 of ECG, edge $(u, v)$ is weighted by how often $u$ and $v$ end up in the same part after 1 step of Louvain. 
Given some partition $\cC$ of $V(G)$ with $u \in C_u$ and $v \in C_v$, write
\[
\mathrm{ecg}\big((u, v), \cC \big) = \begin{cases}
    1 & C_u = C_v\\
    0 & C_u \neq C_v\\
\end{cases}
\]
Then the weight assigned to $(u, v)$ at the end of step~2 of the ECG algorithm is $\sum \mathrm{ecg}\big((u, v), \cC \big)$ where the sum is taken over the $k$ partitions. We consider replacing this weight with a new weight based on a given CAS score. Intuitively, we want $(u, v)$ to receive a higher weight if $u$ is strongly associated with $C_v$ and/or $v$ is strongly associated with $C_u$, even when they are in different parts.
For a $\mathrm{CAS}$ scoring function $f$, we propose the following two options for weighting the edges.
\begin{align*}
f_{or}\big((u, v), \cC\big) 
&= 
f(u, C_v) + f(v, C_u) - f(u, C_v) \cdot f(v, C_u) \,. \\
f_{and}\big((u, v), \cC\big) &= f(u, C_v) \cdot f(v, C_u).
\end{align*}

\begin{figure}[ht]
    \centering
    \includegraphics[width=0.45\linewidth]{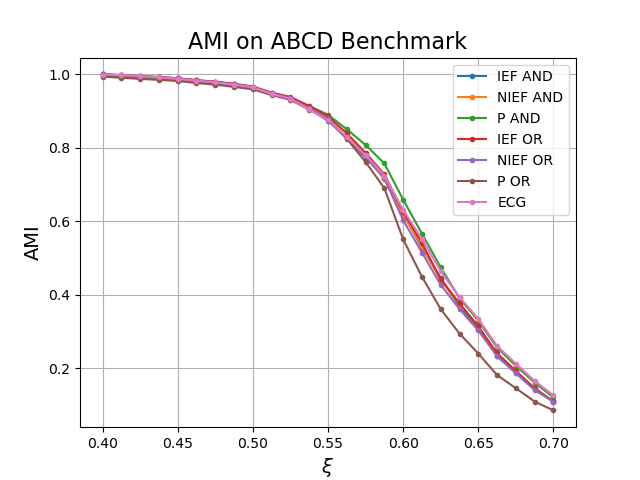}
    \includegraphics[width=0.45\linewidth]{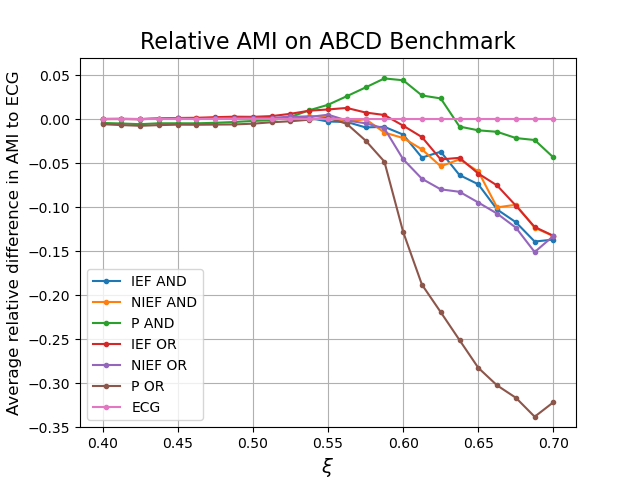}
    \caption{(Left) Average AMI using each CAS and weight functions and the average relative change compared to the base ECG method (Right).
    Any option performs similarly to ECG, with $P_{and}$ performing slightly better on graphs with $\xi$ between $0.55$ and $0.65$.
    For each value of $\xi$, 50 ABCD graphs were generated.}
    \label{fig:casecg}
\end{figure}

We aim to find a CAS score $f$ and a weighting scheme $f_{or}$ or $f_{and}$ that improves ECG. Thus, we test the performance of the six combinations of scores and weighting schemes to see if any can improve ECG's ability to recover ground-truth communities. We perform this test on ABCD graphs with 10,000 nodes, a minimum degree of 5, a minimum community size of 50, and varying levels of noise. Each clustering uses $k = 16$ runs in the ensemble step and obtains the final clustering from running the Leiden algorithm on the weighted graph. The results of this experiment are presented in Figure~\ref{fig:casecg}. We find that each of the six modifications provides comparable results to the base ECG algorithm for $\xi < 0.55$. Although five out of the six configurations seem to yield comparable or worse results than ECG, we find that using $\mathrm{P}$ as the CAS score and $\mathrm{P}_{and}$ as the edge weighting function results in a substantial increase in the Adjusted Mutual Information (AMI) score between noise levels $\xi = 0.55$ and $\xi = 0.65$, with a peak increase of about $5\%$. As can be seen in the left plot, this range of noise levels corresponds to the ``critical'' region where improvements are significant; with lower noise values, most algorithms will yield good results, while for higher noise values, the resulting AMI scores are very low no matter which algorithm is used. 

\subsection{Outlier detection}
\label{subsec:outliers}

\begin{figure}[h!]
    \centering
    \includegraphics[width=5cm]{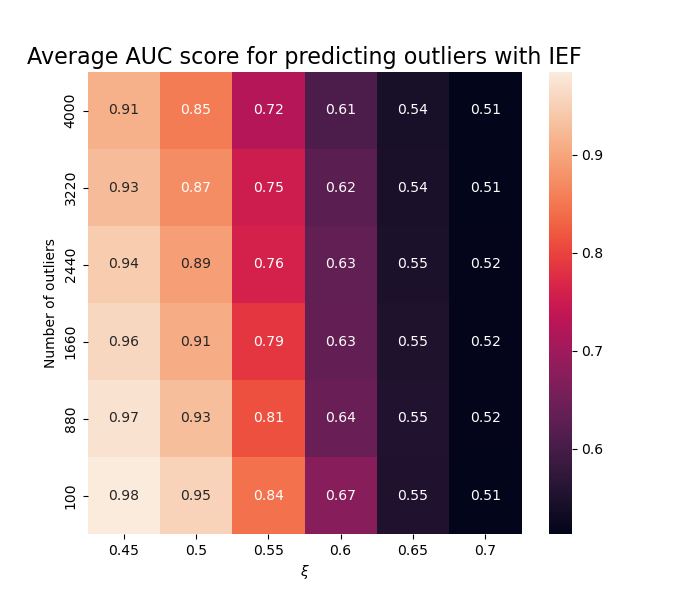}
    \hspace{-.33cm}
    \includegraphics[width=5cm]{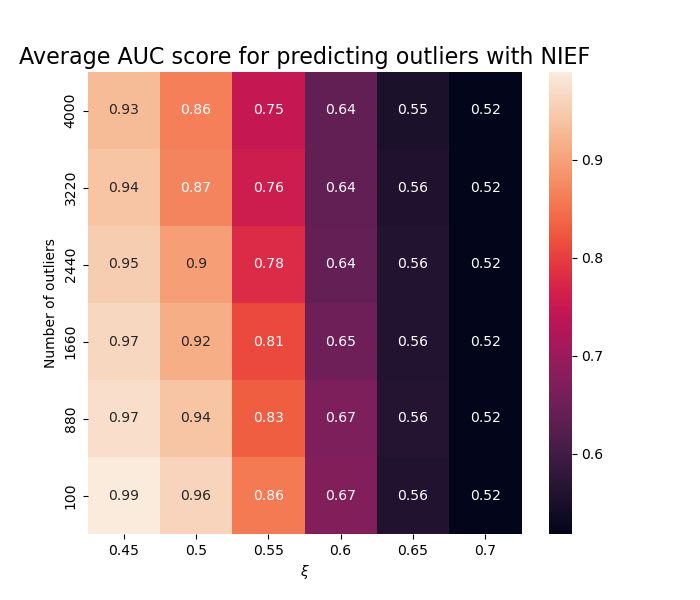}
    \hspace{-.33cm}
    \includegraphics[width=5cm]{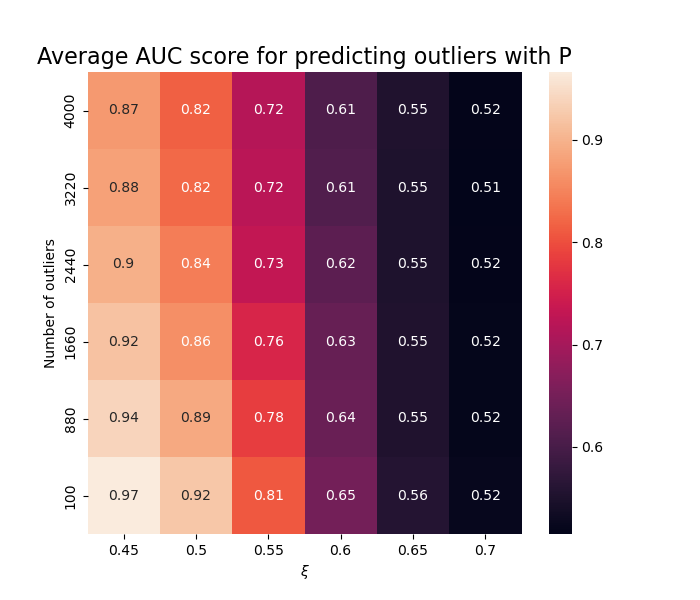}
    \caption{Average AUC score using Leiden followed by CAS-based rankings on 50 ABCD-o graphs with 10,000 nodes.}
    \label{fig:abcdo_roc}
\end{figure}

This section uses a CAS score to post-process a partitioning algorithm to find outliers. Let $G = (V, E)$ be a graph with ground-truth communities and a set of outliers $V_o \subseteq V$, and let $\cC$ be a partition of $V$ found by a detection algorithm. Then, although $\cC$ is not the ground truth assuming $|V_o|>0$, we can still attempt to recover as much of $V_o$ as possible by finding nodes $v$ such that $\mathrm{CAS}(v, C) \approx 0$ for all $C \in \cC$. To test if this heuristic is feasible, we perform the following experiment on the ABCD+o model using each of the 3 CAS scores. First, we use the Leiden algorithm to obtain a partition of the nodes. We then use the CAS score of a node and its suggested community to rank the nodes from most likely to be an outlier (low maximum CAS score) to least likely to be an outlier (high maximum CAS score). Finally, we compute the area under the ROC curve (AUC) for our outlier prediction. For this experiment, we fix $|V| = 10,000$, and we fix the distributions for sampling degrees and community sizes. We test noise parameter $\xi$ varying from $0.45$ to $0.7$, and a number of outliers varying from $100$ to $4,000$ (corresponding to $\%1$ to $\%40$ of the nodes). The results are shown in Figure~\ref{fig:abcdo_roc} and are averaged over 50 independent graphs for each configuration. We observe decreasing performance as the noise level $\xi$ increases and slightly decreasing performance as the number of outliers increases. Each measure performs similarly, with a slight advantage for the NIEF score. These results show that on ABCD+o graphs and using Leiden to find an initial partition, all 3 CAS scores can fairly accurately recover the outliers when $\xi \leq 0.5$, whereas none of the scores can recover the outliers when $\xi \geq 0.6$.

\subsection{Multi-communities}
\label{subsec:multi-community vetices}

Finding overlapping communities in a graph is substantially more difficult than finding outliers, particularly when the graphs are noisy and/or the communities are numerous and small. It can quickly become difficult to distinguish noise edges from within-community edges. Here, we show how to use a CAS score to refine a collection of communities. Let $G = (V, E)$ be a graph, $\cC = \{C_1, \dots, C_k\}$ be a collection of communities in $G$ found by some algorithm, $f$ be a CAS score, and $\tau > 0$ be a threshold. Construct a new collection of communities, $\cC'$, as follows. 
\begin{enumerate}
    \item Initially, $\cC' = \{C_1', \dots, C_k'\}$ is a collection of empty sets. 
    \item For all $v \in V$ and all $C_i \in \cC$, if $f(v, C_i) \geq \tau$ then add $v$ to $C_i'$ in $\cC'$. 
\end{enumerate}

\begin{figure}[ht]
    \centering
    \includegraphics[width=8.15cm]{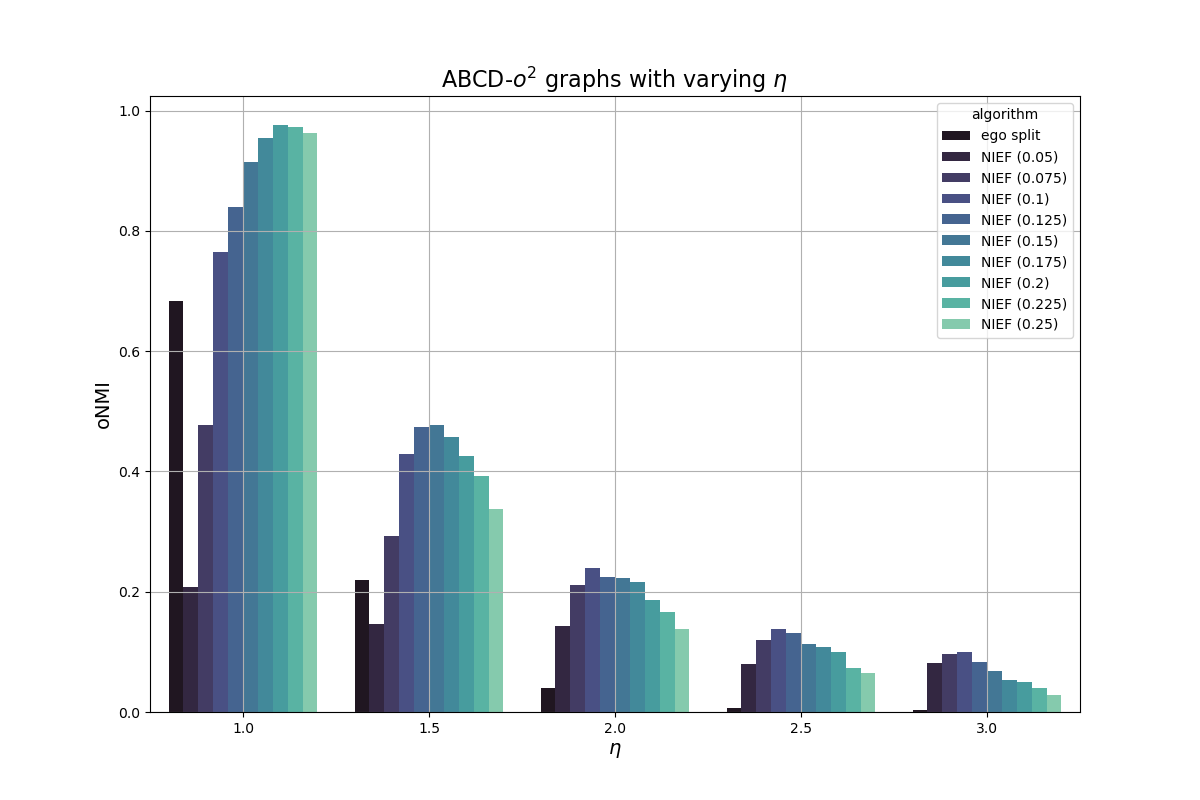}
    \includegraphics[width=8.15cm]{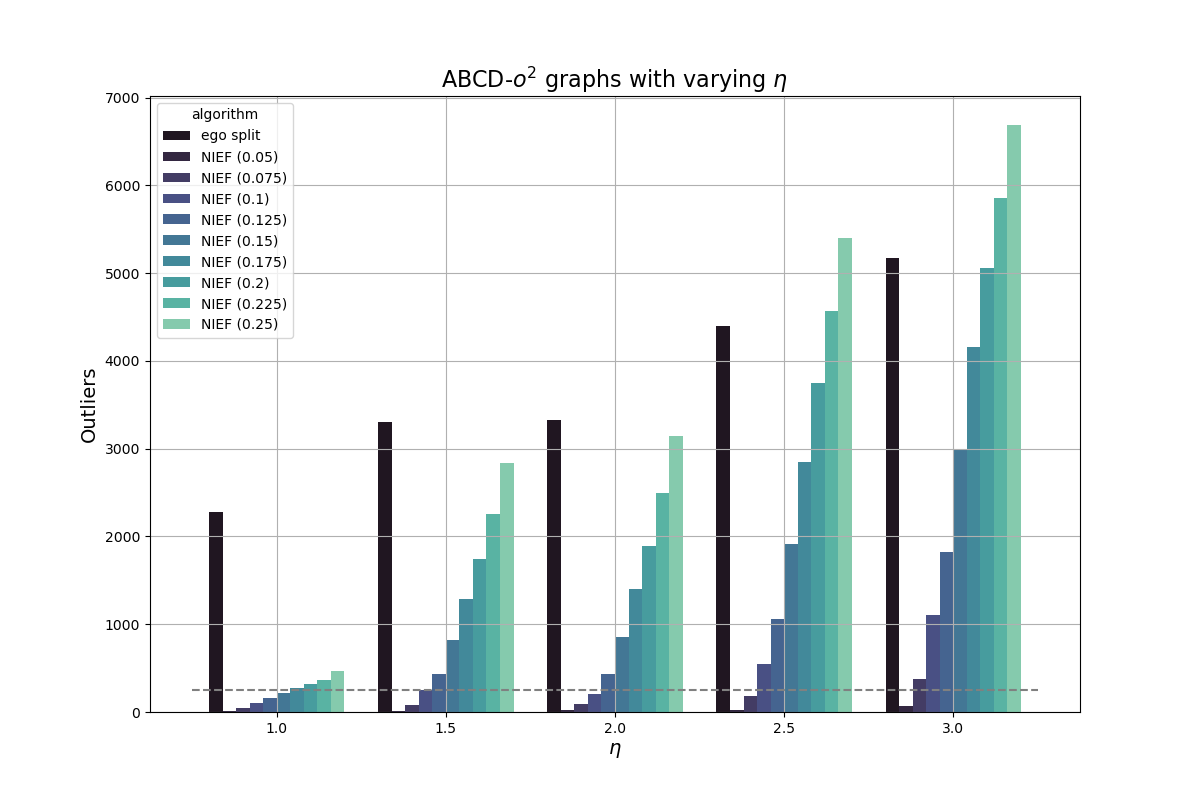}
    \caption{Ego-splitting (darker bars) followed by NIEF with varying threshold values for ABCD+o$^2$ graphs with 10,000 nodes, $\xi = 0.35$, and varying $\eta$ value. We compare each set of communities with the ground truth via the oNMI measure (left), and the number of outlier nodes produced in each case (right).}
    \label{fig:ego_1}
\end{figure}

To test this refinement process, we use the NIEF score to refine communities obtained via the ego-split method (as described in~\cite{kaminski2021mining}) on ABCD+o$^2$ graphs with 10,000 nodes, a fixed noise level $\xi = 0.35$, and varying $\eta$ values. The results are presented in Figure~\ref{fig:ego_1} (left). To measure the quality of predicted communities compared to the ground truth, we use the overlapping Normalized Mutual Information (oNMI) measure: a similarity measure for two collections of subsets $\cX, \cY$ of a set $S$~\cite{mcdaid2011normalized}. For each $\eta \in \{1, 1.5, 2, 2.5, 3\}$, we first find $\cC_{guess}$ via the ego-split method (restricting the minimum community size to 10) and compute $\mathrm{oNMI}(\cC_{guess}, \cC_{true})$. We then obtain $\cC'_{guess}$ using our refinement method with a variety of thresholds $\tau$ and compare the resulting oNMI scores with the original. We find that, while no single value for $\tau$ is clearly the best, choosing $\tau \in [0.075, 0.25]$ yields a refinement $\cC'_{guess}$ that is better than the initial prediction $\cC_{guess}$. This result suggests that, with a well-chosen CAS score and threshold, the refinement process can indeed improve existing detection algorithms. 

In practical applications, since the ground truth is unknown, it is not always clear how to pick a good threshold for the refinement process. We propose a guiding method in Figure \ref{fig:ego_1} (right) where we show the number of outliers obtained for each choice of threshold. The true number of outlier nodes is 250 (shown with a dashed line). While this information is also not likely known beforehand, it gives us some rule of thumb to set the threshold. For example, if we suspect the number of outliers to be small (as it is), then a threshold around 0.1 for NIEF seems like a good choice, possibly slightly higher if we suspect no or little overlap ($\eta \approx 1$) and slightly lower if we suspect lots of overlap. This threshold corresponds to good results in Figure \ref{fig:ego_1} (left).


\subsection{Illustration on a real graph}
\label{subsec:football}

We now illustrate the methods described in Sections~\ref{subsec:wECG}, \ref{subsec:outliers}, and~\ref{subsec:multi-community vetices} using a real-world graph. We consider the college football graph from~\cite{girvan2002community} with corrections to the labels as described in~\cite{evans2010clique}.
The graph has 115 nodes (teams) and 613 edges (games played). After the corrections, there are 12 communities corresponding to football conferences. 
In general, teams play most games within their conference. One of these communities is in fact a group of independent teams which we use as a surrogate for outlier nodes. 

\begin{figure}[ht]
    \centering
    \includegraphics[width=7cm]{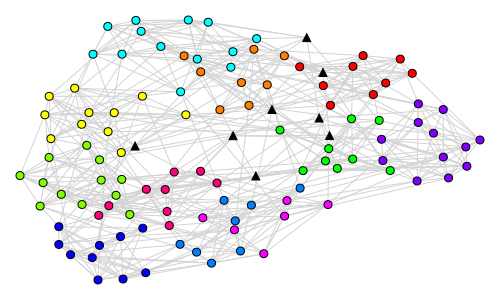}
    \hspace{.5cm}
    \includegraphics[width=7cm]{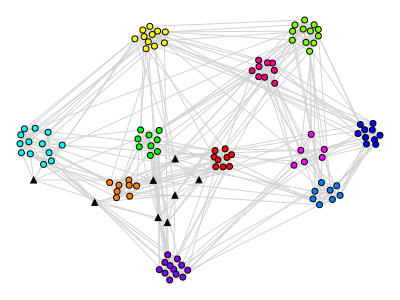}
    \caption{Football graph displayed using a force-directed layout algorithm. The left plot has all unit edge weights, while the right plot has edge weights derived from the CAS-ECG algorithm. Communities are shown in colors, and outlier nodes as black triangles.}
    \label{fig:foot1}
\end{figure}

We first consider the CAS-ECG algorithm from Section~\ref{subsec:wECG}. In Figure~\ref{fig:foot1}, we show the graph using a forced directed layout\footnote{the Fruchterman-Reingold algorithm in Python-igraph}, where the node colours correspond to the different conferences, and the outlier nodes are shown as black triangles (left plot). Running the CAS-ECG algorithm using the $\mathrm{P}_{and}$ weighting scheme, the same force-directed layout algorithm yields the (much nicer) plot on the right. Moreover, we get slightly better communities with CAS-ECG than with Leiden, i.e.\ larger AMI values.

\begin{figure}[ht]
    \centering
    \includegraphics[width=7cm]{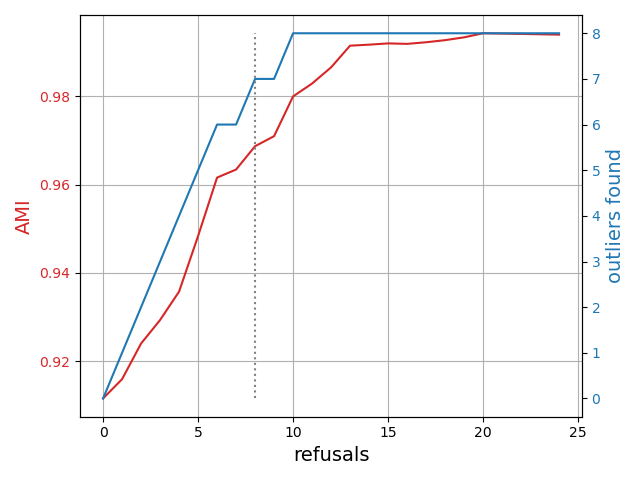}
    \includegraphics[width=7cm]{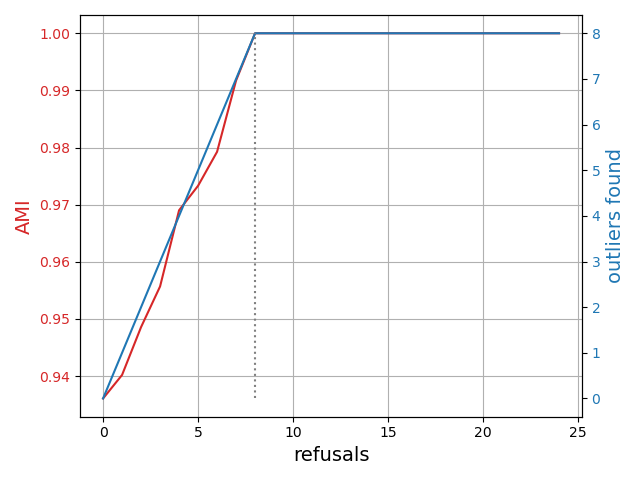}
    \caption{Outlier detection on the football graph respectively using Leiden (left) and the $\mathrm{P}_{and}$ modified ECG (right), followed by node ranking via P score. We show the number of outliers found (blue curves) and the corresponding AMI (red curves) as we iterate through the ranked list.}
    \label{fig:foot2}
\end{figure}

Next, we use the P score to rank nodes as possible outliers as per Section~\ref{subsec:outliers}. The results are presented in Figure~\ref{fig:foot2}. We run two versions of the experiment, the first using Leiden to obtain the initial partition and the second using the $\mathrm{P}_{and}$ version of CAS-ECG. We see that the eight outlier nodes are found in the top 10 (Leiden) or the top 8 (CAS-ECG). Moreover, the CAS-ECG algorithm slightly outperforms Leiden in recovering the ground-truth communities.

\begin{figure}[ht]
    \centering
    \includegraphics[width=6cm]{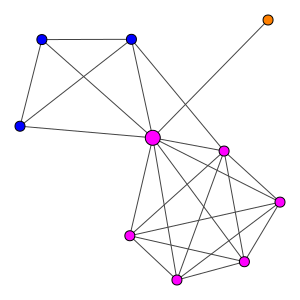}
    \hspace{1cm}
    \includegraphics[width=6cm]{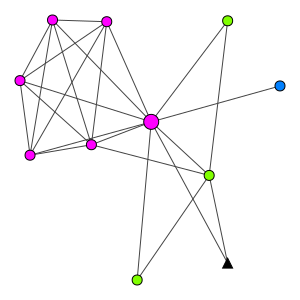}
    \caption{Ego-nets for two nodes (shown with larger circles) having large CAS P-scores toward two communities.}
    \label{fig:foot3}
\end{figure}

Finally, while there is no clear community overlap in this dataset, we look at the most likely node(s) that are part of multiple communities, again using the $\mathrm{P}_{and}$ version of CAS-ECG. In Figure~\ref{fig:foot3}, we show the ego-nets for two nodes with high P scores for two communities. The nodes are shown as larger circles. In both cases, while the nodes are both part of a tight community (magenta), they seem to act as a bridge to other communities.

\section{Conclusion}\label{sec:conclusion}

We presented three community association strength functions, highlighted their similarities and differences, and showed their ability to recover community involvement in a network. We suggested multiple ways to leverage community association strength when detecting communities with overlap and outliers. While the experiments in Section~\ref{sec:post-processing} suggest that CAS scores do a good job at improving community detection, our goal is merely to show that improvement is \textit{possible}. On the one hand, our experiments should be tested on a wide variety of real datasets with ground-truth communities to obtain more conclusive results. On the other hand, there might be stronger CAS scores and/or more clever uses of these scores that outperform what we have presented here. In an upcoming journal version, we will present more experiments on more datasets and delve deeper into the meaning behind the results. Furthermore, we leave it as an open problem to find CAS scores that can outperform the three scores presented here. 

\bibliography{ESCBib}

\begin{thebibliography}{10}

\bibitem{ahn2010link}
Yong-Yeol Ahn, James~P Bagrow, and Sune Lehmann.
\newblock Link communities reveal multiscale complexity in networks.
\newblock {\em nature}, 466(7307):761--764, 2010.

\bibitem{blondel2008fast}
Vincent~D Blondel, Jean-Loup Guillaume, Renaud Lambiotte, and Etienne Lefebvre.
\newblock Fast unfolding of communities in large networks.
\newblock {\em Journal of statistical mechanics: theory and experiment}, 2008(10):P10008, 2008.

\bibitem{calderer2021biological}
Kuijjer~ML. Calderer~G.
\newblock Community detection in large-scale bipartite biological networks.
\newblock {\em Frontiers in Genetics}, 12, 2021.
\newblock \href {https://doi.org/10.3389/fgene.2021.649440} {\path{doi:10.3389/fgene.2021.649440}}.

\bibitem{chung2006complex}
Fan Chung~Graham and Linyuan Lu.
\newblock {\em Complex graphs and networks}.
\newblock Number 107 in CBMS Regional Conference Series in Mathematics. American Mathematical Soc., 2006.

\bibitem{derenyi2005clique}
Imre Der{\'e}nyi, Gergely Palla, and Tam{\'a}s Vicsek.
\newblock Clique percolation in random networks.
\newblock {\em Physical review letters}, 94(16):160202, 2005.

\bibitem{epasto2017ego}
Alessandro Epasto, Silvio Lattanzi, and Renato Paes~Leme.
\newblock Ego-splitting framework: From non-overlapping to overlapping clusters.
\newblock In {\em Proceedings of the 23rd ACM SIGKDD international conference on knowledge discovery and data mining}, pages 145--154, 2017.

\bibitem{evans2010clique}
Tim~S Evans.
\newblock Clique graphs and overlapping communities.
\newblock {\em Journal of Statistical Mechanics: Theory and Experiment}, 2010(12):P12037, 2010.

\bibitem{girvan2002community}
Michelle Girvan and Mark~EJ Newman.
\newblock Community structure in social and biological networks.
\newblock {\em Proceedings of the national academy of sciences}, 99(12):7821--7826, 2002.

\bibitem{han2024social}
Zx. Han, Ll. Shi, and L.~et~al. Liu.
\newblock H-{L}ouvain: Hierarchical {L}ouvain-based community detection in social media data streams.
\newblock {\em Peer-to-Peer Networking and Applications}, 17, 2024.
\newblock \href {https://doi.org/10.1007/s12083-024-01689-9} {\path{doi:10.1007/s12083-024-01689-9}}.

\bibitem{jakatdar2022aoc}
Akhil Jakatdar, Baqiao Liu, Tandy Warnow, and George Chacko.
\newblock {AOC}: Assembling overlapping communities.
\newblock {\em Quantitative Science Studies}, 3(4):1079--1096, 2022.

\bibitem{kaminski2022modularity}
Bogumi{\l} Kami{\'n}ski, Bartosz Pankratz, Pawe{\l} Pra{\l}at, and Fran{\c{c}}ois Th{\'e}berge.
\newblock Modularity of the {ABCD} random graph model with community structure.
\newblock {\em Journal of Complex Networks}, 10(6):cnac050, 2022.
\newblock \href {https://doi.org/10.1093/comnet/cnac050} {\path{doi:10.1093/comnet/cnac050}}.

\bibitem{kaminski2024predicting}
Bogumi{\l} Kami{\'n}ski, Pawe{\l} Pra{\l}at, Fran\c{c}ois Th\'eberge, and Sebastian Zajac.
\newblock Predicting properties of nodes via community-aware features.
\newblock {\em Social Network Analysis and Mining}, 14(1):117, 2024.
\newblock \href {https://doi.org/10.1007/s13278-024-01281-2} {\path{doi:10.1007/s13278-024-01281-2}}.

\bibitem{kaminski2021artificial}
Bogumi{\l} Kami{\'n}ski, Pawe{\l} Pra{\l}at, and Fran{\c{c}}ois Th{\'e}berge.
\newblock Artificial benchmark for community detection ({ABCD})—fast random graph model with community structure.
\newblock {\em Network Science}, pages 1--26, 2021.

\bibitem{kaminski2021mining}
Bogumi{\l} Kami{\'n}ski, Pawe{\l} Pra{\l}at, and Fran{\c{c}}ois Th{\'e}berge.
\newblock {\em Mining Complex Networks}.
\newblock Chapman and Hall/CRC, 2021.

\bibitem{kaminski2023artificial}
Bogumi{\l} Kami{\'n}ski, Pawe{\l} Pra{\l}at, and Fran{\c{c}}ois Th{\'e}berge.
\newblock Artificial benchmark for community detection with outliers (abcd+o).
\newblock {\em Applied Network Science}, 8(1):25, 2023.

\bibitem{kaminski2022hypergraph}
Bogumi{\l} Kami{\'n}ski, Pawe{\l} Pra{\l}at, and Fran{\c{c}}ois Th{\'e}berge.
\newblock Hypergraph artificial benchmark for community detection (h-abcd).
\newblock {\em Journal of Complex Networks}, 11(4):cnad028, 2023.

\bibitem{kaminski2022abcde}
Bogumił Kamiński, Tomasz Olczak, Bartosz Pankratz, Paweł Prałat, and François Théberge.
\newblock Properties and performance of the abcde random graph model with community structure.
\newblock {\em Big Data Research}, 30:100348, 2022.
\newblock URL: \url{https://www.sciencedirect.com/science/article/pii/S2214579622000429}, \href {https://doi.org/10.1016/j.bdr.2022.100348} {\path{doi:10.1016/j.bdr.2022.100348}}.

\bibitem{lancichinetti2008benchmark}
Andrea Lancichinetti, Santo Fortunato, and Filippo Radicchi.
\newblock Benchmark graphs for testing community detection algorithms.
\newblock {\em Physical review E}, 78(4):046110, 2008.

\bibitem{lancichinetti2011statisticallysignificantcommunities}
Andrea Lancichinetti, Filippo Radicchi, José~J. Ramasco, and Santo Fortunato.
\newblock Finding statistically significant communities in networks.
\newblock {\em PLoS ONE}, 6(4):e18961, 2011.
\newblock \href {https://doi.org/10.1371/journal.pone.0018961} {\path{doi:10.1371/journal.pone.0018961}}.

\bibitem{luzar2014community}
Borut Lužar, Zoran Levnajić, Janez Povh, and Matjaž Perc.
\newblock Community structure and the evolution of interdisciplinarity in {S}lovenia's scientific collaboration network.
\newblock {\em PLOS ONE}, 9(4):1--5, 04 2014.
\newblock \href {https://doi.org/10.1371/journal.pone.0094429} {\path{doi:10.1371/journal.pone.0094429}}.

\bibitem{mcdaid2011normalized}
Aaron~F McDaid, Derek Greene, and Neil Hurley.
\newblock Normalized mutual information to evaluate overlapping community finding algorithms.
\newblock {\em arXiv preprint arXiv:1110.2515}, 2011.

\bibitem{newman2004finding}
Mark~EJ Newman and Michelle Girvan.
\newblock Finding and evaluating community structure in networks.
\newblock {\em Physical review E}, 69(2):026113, 2004.

\bibitem{Papadopoulos2012CommunityDI}
Symeon Papadopoulos, Yiannis Kompatsiaris, Athena Vakali, and Ploutarchos Spyridonos.
\newblock Community detection in social media.
\newblock {\em Data Mining and Knowledge Discovery}, 24:515--554, 2012.
\newblock URL: \url{https://api.semanticscholar.org/CorpusID:15719130}.

\bibitem{poulin2018ensemble}
Val{\'e}rie Poulin and Fran{\c{c}}ois Th{\'e}berge.
\newblock Ensemble clustering for graphs.
\newblock In {\em International Conference on Complex Networks and their Applications}, pages 231--243. Springer, 2018.

\bibitem{poulin2019ecgcomparisons}
Valérie Poulin and François Théberge.
\newblock Ensemble clustering for graphs: comparisons and applications.
\newblock {\em Applied Network Science}, 4(4), 2019.
\newblock \href {https://doi.org/10.1007/s41109-019-0162-z} {\path{doi:10.1007/s41109-019-0162-z}}.

\bibitem{rahiminejad2019topological}
S.~Rahiminejad, M.R. Maurya, and S.~Subramaniam.
\newblock Topological and functional comparison of community detection algorithms in biological networks.
\newblock {\em BMC Bioinformatics}, 20, 2019.
\newblock \href {https://doi.org/10.1186/s12859-019-2746-0} {\path{doi:10.1186/s12859-019-2746-0}}.

\bibitem{traag2019louvain}
Vincent~A Traag, Ludo Waltman, and Nees~Jan Van~Eck.
\newblock From {L}ouvain to {L}eiden: guaranteeing well-connected communities.
\newblock {\em Scientific reports}, 9(1):5233, 2019.

\end{thebibliography}

\end{document}